\documentclass[aps,prl,reprint,groupedaddress]{revtex4-2}

\usepackage{graphicx}
\usepackage{xcolor}
\usepackage{amsmath, amssymb}

\begin{document}


\title{Addressing the sign-problem in Euclidean path integrals with radial basis function neural networks}


\author{G\'abor Balassa}
\affiliation{Department of Physics, Yonsei University, Seoul 03722, Korea}


\date{\today}

\begin{abstract}
Solving interacting field theories at finite densities remains a numerically and conceptually challenging task, even with modern computational capabilities. In this paper, we propose a novel approach based on an expansion of the Euclidean path integrals using radial basis function neural networks, which allows the calculation of observables at finite densities and overcomes the sign problem in a numerically very efficient manner. The method is applied to an interacting complex scalar field theory at finite chemical potential in 3+1 dimensions, which exhibits both the sign problem and the silver blaze phenomenon, similar to QCD. The critical chemical potential at which phase transition occurs is estimated to be $\mu_c=1.17 \pm 0.018$, and the silver blaze problem is accurately described below $\mu_c$. 
\end{abstract}


\maketitle


\section{Introduction}
Addressing the nonperturbative region of quantum field theories requires numerical techniques such as lattice Monte Carlo methods, which need a probabilistic interpretation of the corresponding Euclidean path integrals. Introducing imaginary parts into the path integrals \cite{1,2} brings in a series of technical challenges (like the famous sign problem) that are very hard to overcome even with current techniques. These challenges hinder our ability to fully comprehend high-density, low-temperature systems found in e.g., astrophysical environments like neutron stars. Several methods, such as reweighting \cite{R1} or the use of imaginary chemical potentials \cite{R2}, can address this issue, each having limited success within specific temperature and density regimes \cite{R3,R4}. Stochastic quantization via complex Langevin dynamics shows promise in addressing the sign problem, though convergence and performance issues are still present for more complex systems \cite{3,4,5}.
Developing novel, conceptually distinct models to describe systems of this kind is an important task, where neural networks may offer a promising new approach \cite{7,8,9,NN1}. In this paper we will apply a radial basis function type neural network model to approximate the corresponding Euclidean path integrals that contain nonlinear interactions and/or imaginary terms, thus making it a good candidate to solve previously numerically unmanageable problems in both particle \cite{10} and solid-state physics \cite{11}.

To demonstrate the method, we will consider the interacting complex scalar field theory at finite densities, in which case the corresponding action integral receives an imaginary part due to the chemical potential that couples to the time component of the conserved current that is associated with the global U(1) symmetry \cite{12}. There are two phenomena that we will address in this work. The first is the Bose condensation \cite{13} at a critical chemical potential $\mu_c$ corresponding to a phase transition, while the second is the so-called silver blaze phenomenon \cite{14}, which states that at zero temperatures, the observables, such as the number density $\langle n \rangle$ should be independent of $\mu$ for small chemical potentials. The silver blaze problem is a highly nontrivial aspect of such finite density systems, because it relies on an exact cancellation of the different terms at each $\mu<\mu_c$. If a method is able to describe the silver blaze problem and the phase transition at finite chemical potentials, it is a very good sign that it could be applied to other interesting phenomena as well.

First, the RBF method is briefly described in parallel with the description of the underlying system (the interacting complex scalar field) that it will be applied to. Then, after introducing the model, the method is used to calculate the critical chemical potential that corresponds to Bose condensation at finite densities. Finally, the silver blaze region is addressed through the determination of the number density at small chemical potentials. 
At the end, numerical considerations, time complexity analysis, and future possibilities and extensions of the RBF model are briefly discussed.

\section{Radial basis function expansion of the path integral}
\label{sec:1}
In this section, the RBF expansion of the Euclidean path integral formalism will be discussed in parallel with the description of the interacting $\phi^4$ theory.  The continuum action for the self-interacting complex scalar field in 3+1 dimensions can be given as:
\begin{eqnarray}
\label{eq:1}
S = \int d^4 x |\partial_{\mu} \phi|^2 &+& (m^2-\mu^2)|\phi|^2 + \nonumber \\ 
&&\mu(\phi^* \partial_4\phi-\phi\partial_4\phi^*)+\lambda(\phi^*\phi)^2, 
\end{eqnarray}
where $\phi$ is a complex scalar field, $\mu$ is the chemical potential, while $m$ is the bare mass, and $\lambda$ is the bare coupling parameter. After expressing the complex field as two real fields, $\phi = \frac{1}{\sqrt{2}}(\phi_1 + i\phi_2)$, the discretized action on an $N^4$ lattice with unit lattice spacing (a = 1) takes the form:
\begin{equation}
\label{eq:2}
S=\sum_x S_{0,x}(\mu)+S_{1,x}(m,\lambda),
\end{equation}
where $S_{0,x}(\mu)$ contains the cross terms that come from the field derivatives and is defined as:
\begin{eqnarray}
\label{eq:3}
&&S_{0,x}(\mu) =  4 \sum_{a=1,2}\phi_{a,x}^2  
-\sum_{i=1}^3 \sum_{a=1,2}\phi_{a,x}\phi_{a,x+\hat{i}} -  \\
&&\cosh(\mu)\!\!\!\sum_{a=1,2}\!\! \!\phi_{a,x}\phi_{a,x+\hat{4}} \! +\!
 i\sinh(\mu)\Big[\phi_{1,x}\phi_{2,x+\hat{4}} \!-\! 
 \phi_{2,x}\phi_{1,x+\hat{4}}\Big], \nonumber
\end{eqnarray}
while $S_{1,x}(m,\lambda)$ is given by:
\begin{equation}
\label{eq:4}
S_{1,x}(m,\lambda) = \frac{m^2}{2}\sum_{a=1,2}\phi_{a,x}^2 +\frac{\lambda}{4}\Big(\sum_{a=1,2} \phi_{a,x}^2\Big)^2.
\end{equation}
The action is now written in terms of two fields, and the partition function can be calculated as $Z=\int D\phi_1 D\phi_2 e^{-S}$. It is evident that the full theory consists of an imaginary term in the action, thus making the usual probabilistic interpretation and importance sampling impossible. 

The RBF method starts from the exponentialized action, where the nonlinear interaction terms are expressed as a linear combination of Gaussian kernels \cite{15,16} as:
\begin{equation}
\label{eq:5}
e^{-S_{1,x}(m,\lambda)}\approx \sum_{k=1}^Ka_k\,e^{-A\sum_{a}(\phi_{a,x}-c_{a,k})^2  },
\end{equation}
where $K$ is the number of kernels, $a_k$ are the weights, $A$ is the width, and $c_{1,k}$ and $c_{2,k}$ are the centers of the radial basis functions. In \cite{RBF} a very detailed explanation is given of the model and its application to field theoretical problems at zero chemical potentials. Here, we will extend the method with the necessary steps that are needed to solve the finite density problem.

\begin{figure}
\centering
\includegraphics[width=\columnwidth]{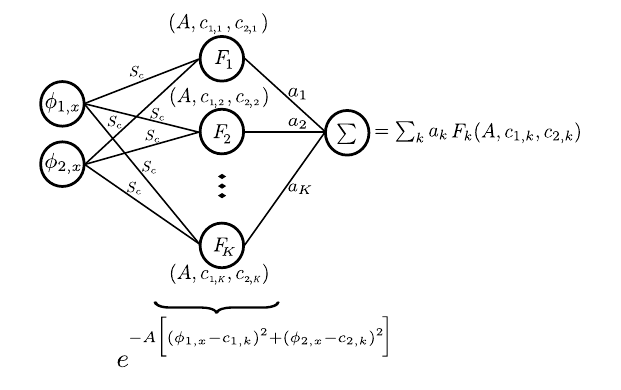}
\caption{General structure of the RBF network with two inputs $\phi_{1,x}$, $\phi_{2,x}$, and $K$ activation functions in the hidden layer. The output corresponds to the weighted sum of the $F_1,F_2,...,F_K$ Gaussian activation functions that have a constant $A$ width parameter and k-dependent $c_{1,k}$, $c_{2,k}$ centers. The $S_c$ weights acts as a scaling parameter of the fields, and is set to $S_c=1$. \label{fig:RBF}}
\end{figure}

According to Eq.~\ref{eq:5}, the corresponding RBF network is a multiple-input-single-output (MISO) system that is defined at each lattice with inputs $\phi_{1,x}$ and $\phi_{2,x}$ and an output that aims to approximate $e^{-S_{1,x}(m,\lambda)}$. Its general structure can be followed in Fig.~\ref{fig:RBF}, where the first layer consists of the inputs, the middle (hidden) layer consists of $K$ number of nonlinear radial activation functions $F_1$, $F_2$, ..., $F_K$, while the final (linear) layer corresponds to a weighted sum of the outputs of the hidden layer with weight factors $a_1$, $a_2$, ..., $a_K$.
The $S_c$ parameters that are shown between the input and the hidden layers are correspond to a possible scaling of the inputs as $\phi \rightarrow S_c \phi$, which could be important if one would want to constraint the inputs to a predefined interval. In this work $S_c$ is always set to $1$, thus it will not be shown explicitly in the equations.
In the hidden layer each nonlinear activation function is a Gaussian that is parametrized by its centers $(c_{1,k},c_{2,k})$, and its width $(A)$.  

Regarding the RBF network, a few important notes have to be made here. First, the width of the kernels ($A$) can be kept constant, as it is always possible to train the network to describe a nonlinear functional form by choosing some corresponding centers, weights, and a possible scaling of the fields. Second, the centers $(c_{1,k},c_{2,k})$ have to be chosen carefully to satisfy a few criteria, which will be defined more precisely later. In general, a good parametrization should include a symmetric center distribution (around zero) that covers the range of the function, with the addition that the RBF approximation should decay very fast outside the range where $e^{-S_{1,x}}$ does not have dominant contributions. The former condition comes from the momentum space behavior of the RBF expansion, while the latter makes sure that one does not have additional contributions when integrating out the path integral. 

The third remark concerns the overall structure of the neural network used in a lattice setting. In general it would be possible to train a global network to describe the system, not just at specific lattice coordinates, but at a finite space-time interval as well. This, however, would require a very careful construction of training and validation samples that are able to mimic all the possible field configurations, e.g., instantons, that could give large contributions to the path integral. In \cite{NN1} this route has been taken and applied to quantum mechanics, where the training samples were generated by using piecewise cubic Hermite polynomials, and the continuum system was estimated in predefined time intervals. In quantum field theory the problem is more nuanced, and it is not a straightforward task to determine the intervals, frequency components, and amplitudes of the training samples that need to be used to accurately describe the dominant field configurations that are necessary to achieve a good generalization. By using a discretized lattice setting, we only need to consider one space-time point at a time, thus, we can overcome this problem, however, it has to be noted that other problems will inadvertently arise that will be shown later.

\begin{figure}
\centering
\includegraphics[width=\columnwidth]{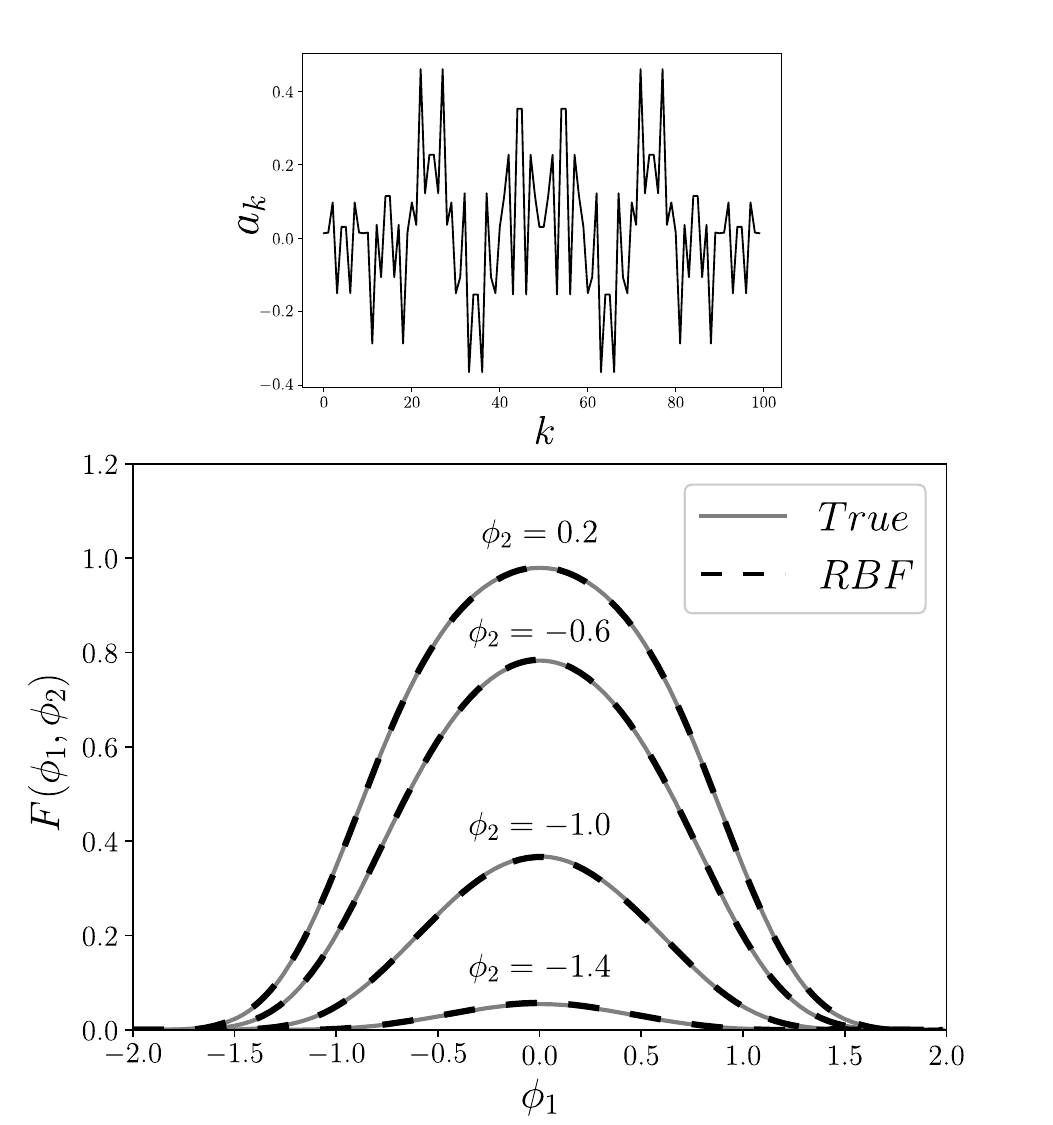}
\caption{Comparison of the true values using $m=1$, and $\lambda=2$, and RBF approximations with  ($A=3$,$c_k \in [-1,1] \times [-1,1]$, $K=10^2$) RBF parameters for fixed $\phi_2$ values. The obtained $a_k$ parameters can be seen on the top figure for the $K=10^2$ basis functions.  \label{fig:RBFtest}}
\end{figure}

Next, a simple example will be shown in the case when the mass is set to $m=1$, and the coupling is defined as $\lambda=2$ to show one possible training procedure of the RBF network. Putting these values into $S_{1,x}(m,\lambda)$, the RBF network can be trained to approximate its values on a finite operating range of $\phi_{1,x}$ and $\phi_{2,x}$. Due to the exponential damping in $e^{-S_{1,x}(m,\lambda)}$, it is an easy task to determine the necessary range for the centers and widths of the Gaussian basis functions. As $e^{-S_{1,x}(m,\lambda)}$ with this parametrization only gives non-negligible values in the range between $\phi_{1,x} \in [-2,2]$ and $\phi_{2,x} \in [-2,2]$ the centers are set to cover a grid between $(c_{1,k},c_{2,k}) \in [-1,1] \times [-1,1]$ with $K=10^2$ centers distributed at equal $\Delta c=(1+1)/(10-1)$ distances in both directions. The width of each Gaussian is set to $A=3$, which is a good choice for these parameterizations. In general, the first step of the training procedure is always the determination of the centers and the widths, and their values depend on the approximable functions. If the function is heavily oscillating, a larger $A$ value could be necessary to be able to capture the high-frequency parts as well. In practice the number of kernels, the distribution of the centers, and the value of the width parameter need to be determined in a way so that the training of the $a_k$ weights can converge to a satisfying value, and the approximable function can be described with good accuracy. To determine the $a_k$ weight parameters, $N_T=1000$ training points are generated randomly in the range of $(\phi_{1,x},\phi_{2,x}) \in [-2.5,2.5] \times [-2.5,2.5]$ and the optimization problem is solved in the least squares sense by first constructing the coefficient matrix $H(c_k,A) \in \mathbb{R}^{N_T \times K}$ and then solving the optimization problem described as follows:
\begin{equation}
\hat{a} = \arg\min_{a} \; \|H(c_k,A) a - y\|^2,
\label{eq:opt}
\end{equation}
where $a=(a_1,a_2,...,a_K)$ is the sought weight parameter vector, and $y=(y_1,y_2,...,y_{N_T})$ is the generated output vector, while the $H(c_k,A)$ coefficient matrix depends on the width parameter $A$ and on the centers $c_k=(c_{1,k},c_{2,k})$. In general this optimization problem can be solved in many ways, e.g., by gradient descent algorithms, singular value decomposition (SVD), etc. In this case the SVD method is used with a small $r_c=0.001$ regularization parameter that helps prevent instabilities in the numerical method by truncating the very small singular values. In general, as the coefficient matrix $H$ depends on the $A$ and $c_k$ parameters, the full optimization procedure should also consist of the determination of the number of kernels $K$, the widths of the Gaussians $A$, and their centers $c_k$. The easiest way to do this is to generate many different configurations, do the optimization shown in Eq.~\ref{eq:opt}, and then choose the best model that has the lowest mean squared error for a randomized test set. A more sophisticated method to address the complexity of the model is to use, e.g., the Akaike Information Criterion (AIC) \cite{AIC}, which is able to compare the different models in a more standardized way. The comparison between the true values and the RBF approximation for the given problem ($m=1$, $\lambda=2$) with the parameters ($A=3$, $c_k \in [-1,1] \times [-1,1]$, $K=10^2$) can be seen in Fig.~\ref{fig:RBFtest} using a test set in the interval of $(\phi_1,\phi_2) \in [-2,2]\times [-2,2]$ on a uniform grid with a grid resolution of $\Delta =0.02$.

The RBF expansion defined this way corresponds to one network at each lattice site, which, if put back into the original path integral, the partition function can be expressed in the following closed form:
\begin{equation}
\label{eq:6}
Z \approx \int \prod d\phi_{1,x}d\phi_{2,x} \, e^{-S_{0,x}(\mu)} \cdot F[\phi_{a,x}],
\end{equation}
where $F[\phi_{a,x}]$ can be given by the radial basis function expansion as:
\begin{equation}
\label{eq:7}
F[\phi_{a,x}]=\sum_{k=1}^Ka_k\,e^{-A\sum \limits_{a} \phi_{a,x}^2 +2A\sum \limits_{a}c_{a,k}\phi_{a,x} - A\sum \limits_{a}c_{a,k}^2},
\end{equation}
where we have expanded the Gaussian into quadratic, linear, and constant terms in the exponent. In general, $S_{0,x}$ can be written in a quadratic form, thus, in matrix notation for all $x$, we will have $S_{0}(\mu) \rightarrow \frac{1}{2}\phi^T M(\mu) \phi$, where $\phi$ is a vector that contains both $\phi_1$ and $\phi_2$ for all $x$, and $M(\mu)$ is a general $\mu$-dependent quadratic matrix with both diagonal and off-diagonal elements. The sum over all $x$ (encoded in the matrix notation) can be written as a product $\prod_x$, in which case we will have a product of sums $\prod_x \sum_k$ for all the lattice sites. This can be reformulated into a large sum of $K^{N^4}$ Gaussian integrals, related to all the possible combinations of the Gaussian kernels. Using this notation, the partition function can be written simply as:
\begin{eqnarray}
\label{eq:8}
Z \approx \!\!\!\!\!\!\!\!\! 
\sum \limits_{k \in \{1,...,K \}^{N^4}} \!\!\!\!\!\!\hat{a}_k\int \mathcal{D}\phi \;e^{-\frac{1}{2}\phi^T W \phi  + 2A(\hat{c}_k^T\phi)-A(1^T\hat{c}_k^2)},
\end{eqnarray}
where we have a sum over all the possible combinations between the terms of the RBF expansion over all the lattice sites, with the notation $k =\{k_x\}_{x=0,1,...N^4}$, where each $k_x \in \{1,...,K \}$. Moreover, we have defined the new matrix $W=[M(\mu)-2\hat{A}]$, where $\hat{A}$ is a diagonal matrix with $A$ in its diagonals, $1^T$ is the $2N^4$-long unit vector ($N^4$ term for each $a=1,2$), $\mathcal{D}\phi = \prod_x d\phi_{1,x}d\phi_{2,x}$ is the integral measure, $\hat{a}_k=\prod_x a_{k_x}$ is the product of all weight factors for a specific combination of the elements $k_x$, while $\hat{c}_k$ is a $2N^4$-long vector containing specific combinations of the corresponding Gaussian centers.

The coordinate space form in itself is not too convenient to use due to the nondiagonal terms. To overcome this issue, let us go into momentum space by applying a $U$ similarity transformation that diagonalizes the $W=[M(\mu)-2\hat{A}]$ matrix in $(\widetilde{\Phi}_{1,p}$, $\widetilde{\Phi}_{2,p})$ space, thus generating a block diagonal matrix with $2\times 2$ submatrices in its diagonal elements. The transformation of the fields reads as $\widetilde{\Phi}=U^T\phi$, where $\widetilde{\Phi}$ is the momentum space vector of the fields $\widetilde{\Phi}_{a,p}$ at momenta $p_i=2 \pi n_i/N$, with $-N/2<n_i\leq N/2 $. The partition function after transforming into momentum space can be written as:
\begin{equation}
\label{eq:9}
Z \approx \!\!\!\!\!\!\!\!\!\!\!\!\sum \limits_{k \in \{1,...,K \}^{N^4}} \!\!\!\!\!\!\! \hat{a}_k\int \!\! \mathcal{D} \widetilde{\Phi} \; e^{-\frac{1}{2} \widetilde{\Phi}^{T}\!(U^T \! WU)\widetilde{\Phi} + 2A(U^T \!\hat{c}_k)\!^T\widetilde{\Phi} -A(1\!^T \!\hat{c}_k)  },
\end{equation} 
where $\widetilde{\Phi}=[\widetilde{\Phi}_{1,p_1},\widetilde{\Phi}_{2,p_1},...,\widetilde{\Phi}_{1,p_j},\widetilde{\Phi}_{2,p_j},...]$ vector consisting of all the transformed fields, while the $\widetilde{W}=U^T W U$ diagonalized matrix.

The full $\widetilde{W}$ matrix is built up by these $2\times2$ submatrices in its diagonal. 
To make the RBF expansion more useful, we would like to express the momentum space path integral in the following factorized form:
\begin{equation}
\label{eq:10}
Z \rightarrow \prod_p  \sum_{k=1}^K Q_k \Big[ \widetilde{\Phi}_{a,p},\mu,A,c_{a,k},a_k\Big],
\end{equation}
which means each momentum should have a separate RBF network contribution. In general this factorization is not possible due to the mixing of the $\hat{c_k}$ terms through the $(U^T\hat{c}_k)$ transformation, however, by a careful selection of the parameters of the RBF network, and due to the special functional form after integration, it is possible to approximate $\ln Z$ with very good accuracy. To quantify this, let us compare the logarithm of the integral of Eq.~\ref{eq:9} with and without the $U^T$ transformation in the $2A(U^T\hat{c}_k)\widetilde{\Phi}$ linear shift terms. First, after integrating out the partition function, we get the following closed-form expression in momentum space:
\begin{eqnarray}
\label{eq:11}
Z &\approx& \!\!\!\!\!\!\!\!\!\sum \limits_{k \in \{1,...,K \}^{N^4}} \!\!\!\hat{a}_k \prod_p \frac{2 \pi} {\sqrt{\mathcal{A}_p^2+\mathcal{B}_p^2}} \exp\Bigg[ 
- A(c_{1,k_p}^2 + c_{2,k_p}^2) + \nonumber \\
&& \quad \quad\frac{2A^2}{\mathcal{A}_p^2 + \mathcal{B}_p^2} \Big\{ 
\mathcal{A}_p \left[(U^T\hat{c}_{1,k})^2_p + (U^T\hat{c}_{2,k})^2_p\right] 
\Big\} \Bigg],
\end{eqnarray}
where the $\mathcal{A}_p$ and $\mathcal{B}_p$ parameters are coming from the diagonalization of the $W$ matrix and can be written as:
\begin{eqnarray}
\label{eq:12}
&&\mathcal{A}_p = -2A + 4\sum_{i=1}^3\sin^2\left(\frac{p_i}{2}\right) + 2\Big[1-\cosh(\mu)\cos(p_4)\Big], \nonumber \\
&&\quad\quad \quad\quad \quad\quad\mathcal{B}_p = 2\sinh(\mu)\sin(p_4).
\end{eqnarray}
The notation $\hat{c}_{a,k}$ in Eq.~\ref{eq:11} represents a vector that is created by a specific combination of the $c_k$ RBF centers e.g. $\hat{c}_{a,k}=[c_{a,1},c_{a,3},c_{a,3},...,c_{a,6}]$, that is related to the $k$'th combination from all the possible $K^{N^4}$ combinations. According to this, $c_{a,k_p}$ is one element from this vector that corresponds to the momentum $p$ (i.e., multiplies $\widetilde{\Phi}_{a,p}$).

To examine the differences that the $U$ transformation makes in the results, we set $a_k=1$ $(\forall k)$ and compare the general expression in Eq.~\ref{eq:11} with the same form, but instead of $(U^T\hat{c}_{a,k})_p$, we use $c_{a,k_p}$ in the 2nd line. We will call the latter $Z_{0}$, and the former $Z_U$, and compare $\ln Z_{0}$ and $\ln Z_{U}$ through their averaged relative errors for different $c_k$ distributions defined as:
\begin{equation}
\label{eq:rel}
E_R(N_c) = \frac{1}{M} \sum_{i=1}^M \frac{| \ln Z_{U,i}(N_c) - \ln Z_{0,i}(N_c) | }{| \ln Z_{U,i}(N_c)|},
\end{equation}
where $M$ is the number of samples we average to, while $N_c$ is the number of combinations that is considered in the sum in Eq.~\ref{eq:11}.
As it is infeasible to include all possible combinations in the sum when $N^4$ is large, we have to use a finite sample size $N_c< K^{N^4}$ that will be used to estimate the properties of the errors. It turns out the relative error converges to a finite value after a few hundred samples, which will be apparent from the simulations. 

\begin{figure}
\centering
\includegraphics[width=\columnwidth]{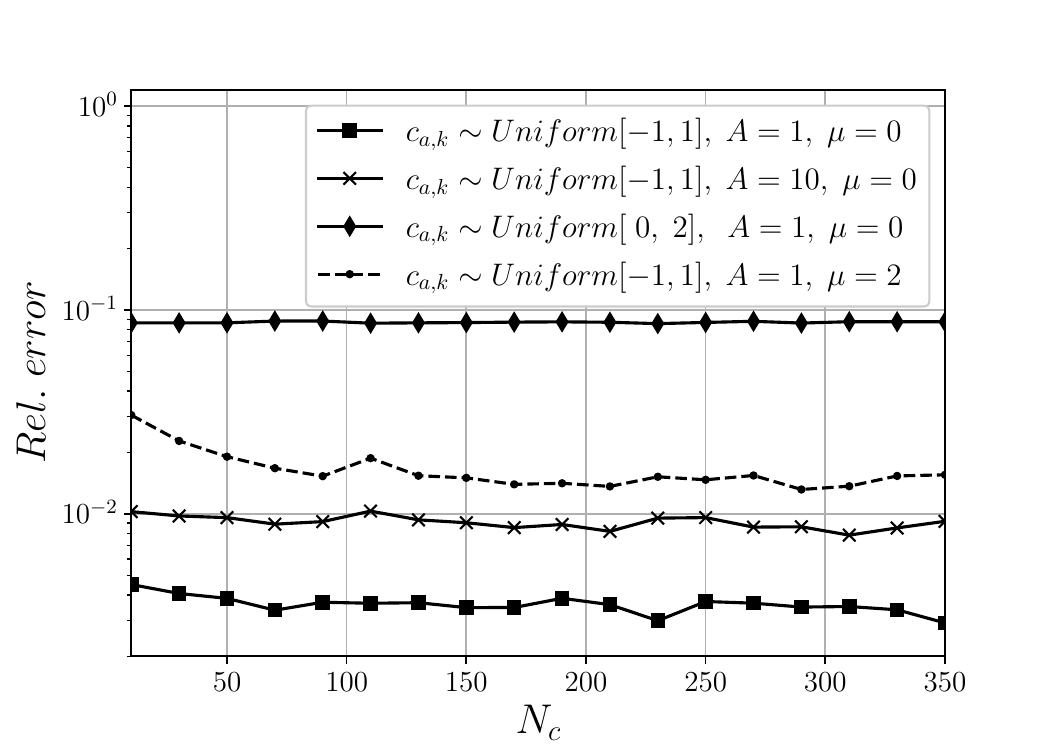}
\caption{Dependence of the relative difference between $\ln Z_{U}$, and $\ln Z_{0}$ on the number of combinations $N_c$ included for different RBF parametrizations and chemical potentials. \label{fig:1}}
\end{figure}

In these simulations we have chosen $K=20$ centers randomly from a uniform distribution between the interval $[a,b]$, and calculated the relative errors for fixed $N_c$ combinations. 
A more detailed description of the error analysis for the free scalar field case can be found in \cite{RBF}. The results for the case of the complex scalar fields at finite chemical potentials can be seen in Fig.~\ref{fig:1} for different $c_k\sim Uniform[a,b]$ uniformly distributed centers, $A$ widths, and $\mu$ chemical potentials.

According to the results, by using symmetric $c_k$ distributions around $0$ and some corresponding $A$ parameters, it seems possible to estimate $\ln Z$ with very good accuracy, using a factorized form in momentum space shown in Eq.~\ref{eq:10}. It is also evident that the error converges to a fixed value after a few hundred samples, and while the error depends on the distribution of the centers, their intervals, and the $A$ and $\mu$ parameters, it stays below a few percent if one chooses symmetric centers around $0$.

\begin{figure}
\centering
\includegraphics[width=\columnwidth]{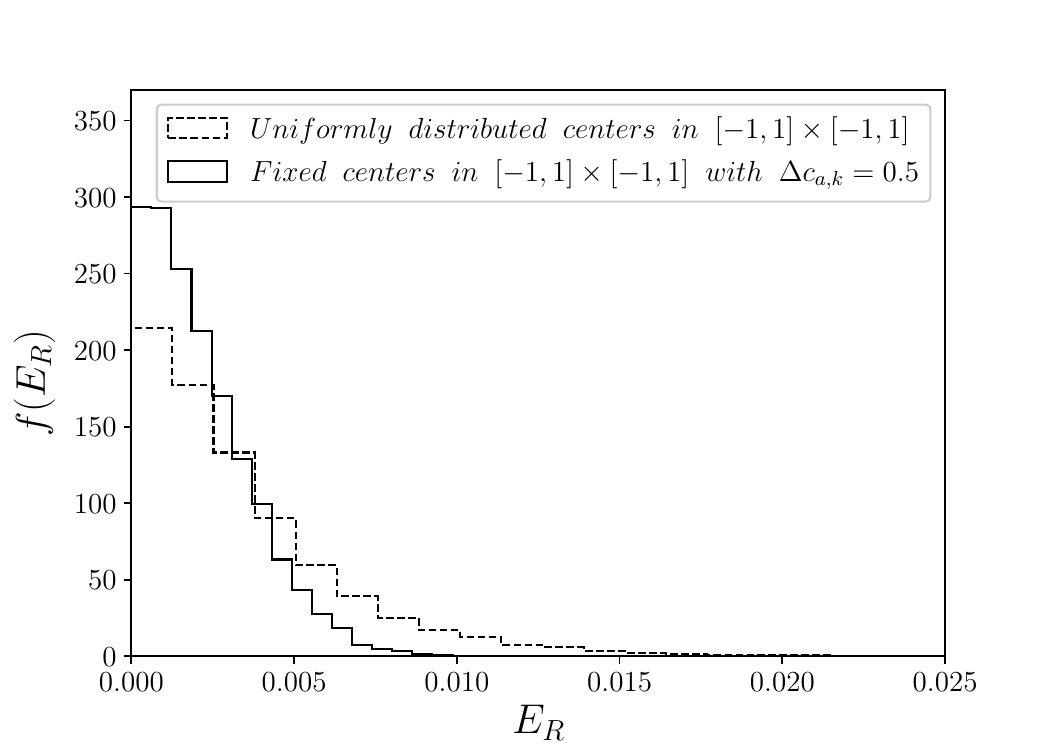}
\caption{Normalized distribution of the relative error for uniformly distributed centers (dashed) and for a fixed center configuration (full) with $K=25$, $A=1$, and $\mu=1$. \label{fig:E}}
\end{figure}

The actual value of the error depends on many factors, e.g., the number of centers, the distribution of centers, the interval, the Gaussian width parameters, etc. In general a fully symmetric center distribution around $0$ appears to be the best choice, which can be seen in Fig.~\ref{fig:E}, where the relative error distribution is compared in the case when in each sample the centers were generated uniformly in the interval $[-1,1] \times [-1,1] $ with $K=25$ number of centers to the case when the centers are fixed at a 2-dimensional grid $[-1,1] \times [-1,1]$ with $\Delta c_{a,k}=0.5$.

From the distributions of the relative errors, it seems that a fully symmetric center distribution on the two-dimensional grid behaves better than the case of uniformly distributed centers. Due to the longer tail and the broader distribution of the uniformly distributed case, the average relative error will be slightly larger than in the other case, however, the difference is not that significant, as the error is still under one percent.

The results suggest that by choosing appropriate parameters, the difference between the full form of Eq.~\ref{eq:11} with and without setting $U$ to $1$, is well controlled, and the full path integral could be written in a factorized form shown in Eq.~\ref{eq:10}, thus, the complexity reduces from $\mathcal{O}(K^{N^4})$ to $\mathcal{O}(K N^4)$, which is a huge improvement and makes the RBF approximation very useful in calculating observables. 

Starting from the momentum space expression in Eq.~\ref{eq:9}, by choosing 'appropriate' $c_k$ and $A$ RBF parameters, the path integral can be approximated as:
\begin{eqnarray}
\label{eq:13}
Z^{rbf} \!\!&=& \!\prod_p\sum_{k=1}^K a_k \int \mathcal{D}\widetilde{\Phi}_{p}\;\text{exp} \Bigg[-\frac{1}{2}\sum_{a,b}\widetilde{\Phi}_{a,-p}\widetilde{W}_{ab,p}\widetilde{\Phi}_{b,p}+ \nonumber \\
&& \quad 2A\sum_a c_{a,k} \widetilde{\Phi}_{a,p}-A\sum_ac_{a,k}^2 \Bigg] 
\end{eqnarray}
where $\mathcal{D}\widetilde{\Phi}_{p}=d\widetilde{\Phi}_{1,p}d\widetilde{\Phi}_{2,p}$, and the complexity has been reduced to $\mathcal{O}(K  N^4)$. Using this formula the generating functional given by $\mathcal{W}=\ln Z$ can be approximated by $\mathcal{W}^{rbf}=\ln Z^{rbf}$ by a few percent accuracy. 
\section{Results}
\label{sec:2}
Using the RBF approximation of the generating functional makes it possible to calculate observables in a very convenient way. In general, due to the Gaussian nature of the expressions, the finite chemical potential should not pose any further problems. Here, we will be interested in two phenomena related to the relativistic Bose gas at finite densities. First, we would like to address the phase transition at some critical chemical potential and determine $\mu_c$ by using the RBF approximation. Secondly, we will address the silver-blaze problem, that is, the $\mu$-independent behavior of the observables at zero temperatures (i.e., large $N_t$) when $\mu<\mu_c$. 
In both cases the system parameters are set to $m=1$, and $\lambda=1$.

There are several methods exist that can help find the phase transition point, e.g., finite size scaling of the binder cumulants \cite{17,18}, susceptibilities, etc. The method that is more suitable to the RBF expansion is the determination of the effective potential \cite{19,20} (i.e., the momentum-independent part of the effective action) defined through the Legendre transform of the generating functional. In practice, we add a constant background source $J^* \phi + J \phi^*$ to the continuum action and define the $S_{1,x}$ term that has to be approximated by an RBF network as:
\begin{equation}
\label{eq:14}
S_{1,x} = \frac{m^2}{2} \!\!\! \sum_{a=1,2}\phi_{a,x}^2 +\frac{\lambda}{4}\Big(\sum_{a=1,2} \phi_{a,x}^2\Big)^2 \!\!+ \!\! \sum_{a=1,2}J_a \phi_{a,x}
\end{equation} 
Due to the $O(2)$ symmetry, we can set $J_2=0$ without loss of generality (i.e., we choose a specific vacuum state on the $\phi_2=0$ axis). The effective potential in this case can be defined as:
\begin{equation}
\label{eq:15}
V_{eff}(\phi_1) \propto \int d \langle \phi_1 \rangle J(\langle \phi_1 \rangle) ,
\end{equation}
where to determine the $J(\langle \phi \rangle)$ function, first we have to calculate $\langle \phi \rangle_{J}$ classical fields at a series of $J$ values so that we can invert it. The classical fields can be calculated by the RBF-expanded generating functional as follows:
\begin{equation}
\label{eq:16}
\langle \phi_1 \rangle_{J} = \frac{\partial \ln Z^{rbf}[J_1]}{\partial J_1} \Big|_{J_1=J}.
 \end{equation}
Due to the dependency of $S_{1,x}$ on the background shift $J_a$ in Eq.~\ref{eq:14}, each $\langle \phi_1 \rangle_J$ calculation needs a training of a separate RBF network that depends on the actual value of the shift in order to be able to obtain the necessary $J(\phi_1)$ function for the effective potential. The training of the RBF networks follows the same procedure that is described in the previous section. The determination of the $K$, $c_k$, $A$, and $a_k$ parameters is done in an automated way by training the networks for each $J$, using many different uniform grids for the centers, defined between $[-4,4] \times [-4,4]$, with different numbers of kernels in the range of $K \in [5^2, 30^2]$. For example, one configuration could consist of a uniform grid of centers in $[-1,1]\times [-1,1]$ with $K=5^2$, while another configuration could be a uniform grid in $[-2,2]\times [-2,2]$ with $K=10^2$. The width parameter is also changed in every network configuration in the range of $A \in [1,15]$. The $a_k$ parameters are determined with each of these parameters by solving the optimization problem in Eq.~\ref{eq:opt}. Then, using randomly generated test data, the best model with the minimal mean squared error is chosen. Using this method, each $J$ could correspond to a different RBF network that, in a mean squared sense, optimally describes the given shifted functional forms.

\begin{figure}
\centering
\includegraphics[width=\columnwidth]{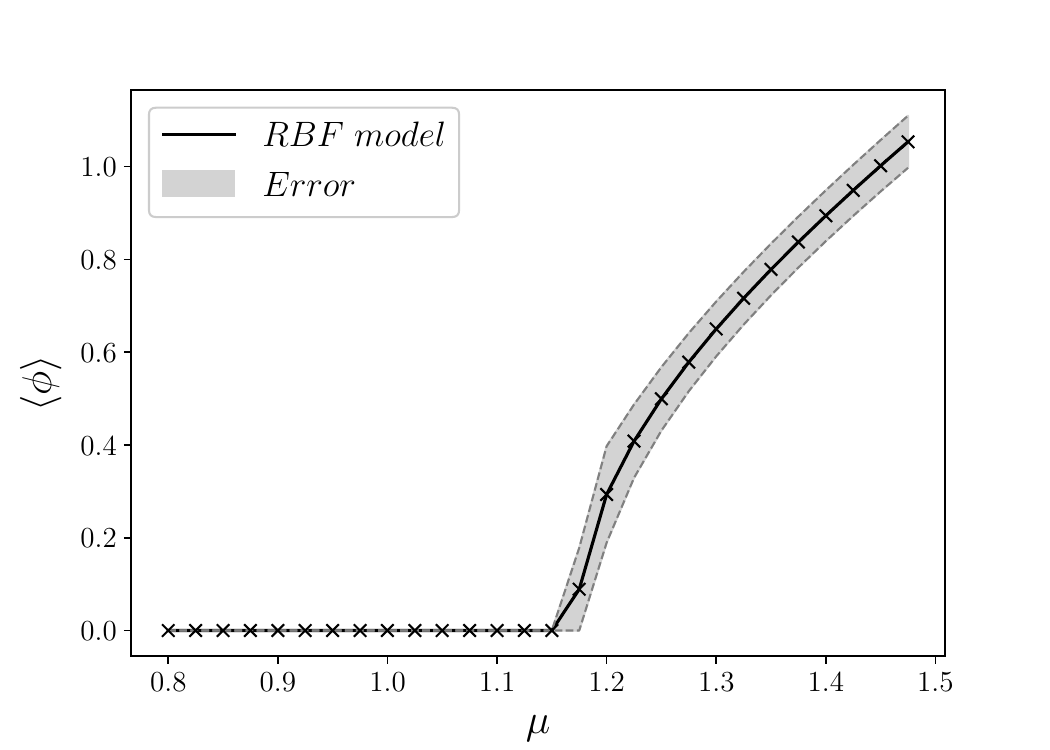}
\caption{Vacuum expectation value of the fields determined from the effective potential. The critical chemical potential is $\mu_c = 1.17 \pm 0.018$. \label{fig:vev}}
\end{figure}

One problem that generally arises is the non-convexity of the effective potential \cite{21,22}, which is a consequence of the Legendre transform that requires convexity of the participating functions. This is related to the Maxwell construction between the saddle points of the effective potentials \cite{23}. The critical chemical potential can be extracted by fitting the $J(\langle \phi_1 \rangle)=a\langle \phi_1 \rangle^3 + b\langle \phi_1 \rangle$ Ansatz for large $J$ shifts, in which case $\mu_c$ can be related to the point where the $b$ parameter becomes negative. This form is related to the $V_{eff}=A(m_R,Z_R)\phi^2 + B(\lambda_R,Z_R)\phi^4$ renormalized potential, with $m_R$, $\lambda_R$, and $Z_R$ renormalized parameters. Using the RBF approximation, $\langle |\phi| \rangle$ can be estimated from the fitted parameters as: $\langle \phi_1 \rangle = \sqrt{|b|/a}$. The results can be seen in Fig.~\ref{fig:vev}, which clearly shows the signs of a phase transition around $\mu_c \approx 1.17 \pm 0.018$. This value is consistent with the one that is obtained using stochastic quantization and complex Langevin dynamics \cite{25}.

\begin{figure}
\centering
\includegraphics[width=\columnwidth]{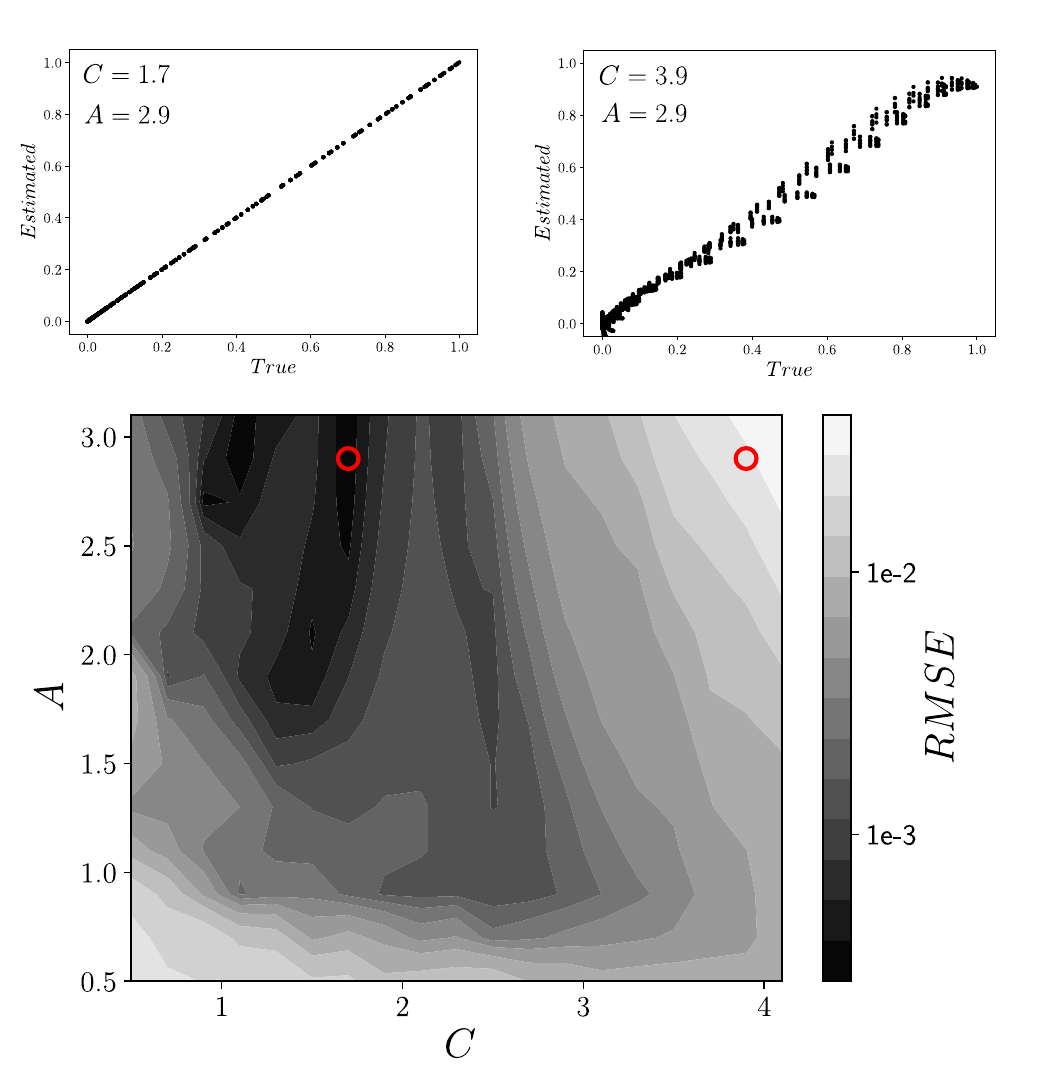}
\caption{RMSE dependence on the width and center parameterizations of the different RBF configurations with $K=10^2$ number of kernel functions. The minimum corresponds to the darkest value at $C=1.7$, and $A=2.9$. The top figures show the scatter plots of the true vs. estimated values for the optimum parametrization (left), and for another case, when the approximation is insufficient to describe the system (right).}
\label{fig:E123}
\end{figure}

\begin{figure}
\centering
\includegraphics[width=\columnwidth]{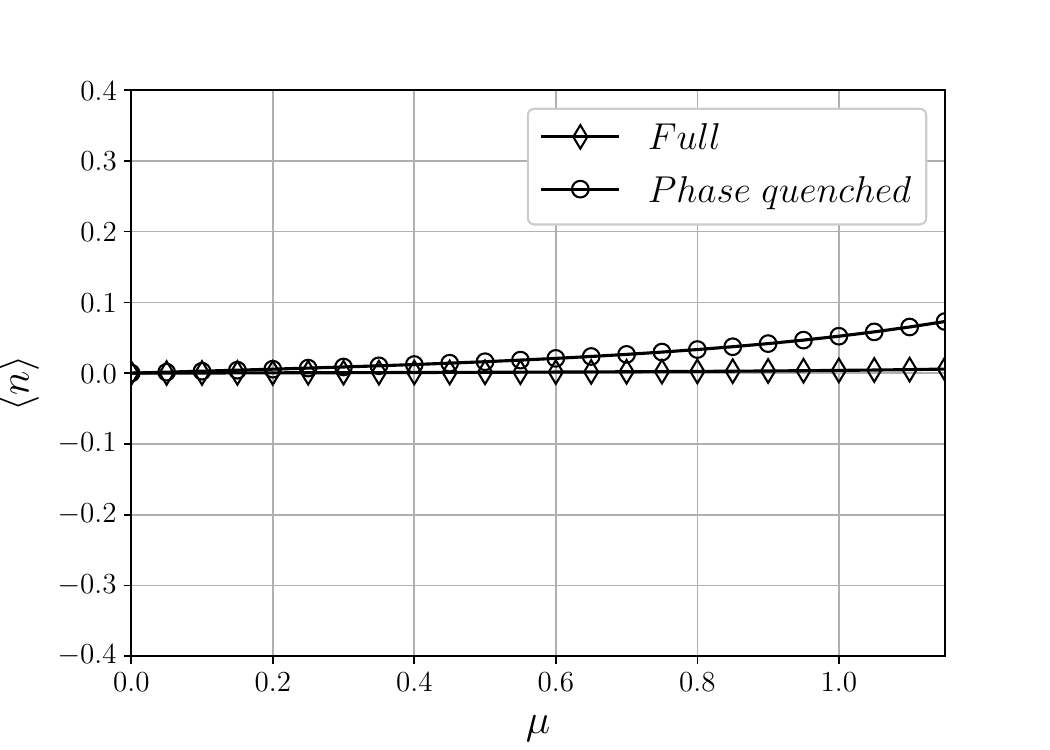}
\caption{Particle number density for the full-, and for the phase quenched theories on a $10^4$ lattice.}
\label{fig:3}
\end{figure}

Next, we will briefly address the silver blaze problem, that is, the $\mu$-independent behavior of the bulk observables at zero temperatures and small chemical potentials (typically under $\mu_c$). Here, we will only consider the particle number density that can be defined through the derivative of the partition function with respect to the chemical potential as:
\begin{equation}
\label{eq:17}
\langle n \rangle = \frac{1}{V}\frac{\partial \ln Z^{rbf}}{\partial \mu}  ,
\end{equation}
where $Z^{rbf}$ is the RBF approximated partition function. The $\mu$ independence is the consequence of severe cancellations due to the imaginary parts in the action, thus, in the phase quenched theory (i.e., when $\mathcal{B}_p=0$ in Eq.~\ref{eq:12}) we do not expect this behavior. 
In contrast to the previous case, when we needed to introduce constant background shifts to the $S_{1,x}$ action in Eq.~\ref{eq:14}, now we only need to approximate one functional form given by the original form of $S_{1,x}$ shown in Eq.~\ref{eq:4}. The optimization procedure follows the previously described method, i.e., optimizing for $a_k$ using a set of $A$, $c_k$, and $K$ parameters, then choosing the model that best describes the test data in the mean squared sense. To show this in more detail, the training procedure has been carried out by assuming a two-dimensional uniform grid of center configuration, then fixing the number of kernels to $K=10^2$, and varying the widths parameters and the center configurations in the range of $A \in [0.5,3.1]$ and $c_k \in [-C,C] \times [-C,C]$, where the parameter $C$ determines the range of the centers and is chosen from $C \in [0.5,4.1]$. The determination of the center configurations this way means that we will have a $10 \times 10$ uniform grid of centers with $\Delta c = 2C/(K-1)$ resolution. To solve the optimization problem for the $a_k$ weight parameters, $2500$ input pairs $(\phi 1,\phi 2)$ were uniformly sampled from the domain $[-2,2]\times [-2,2]$, which was then used for training the corresponding networks. The determination of the best network configuration has been done by calculating the corresponding root mean squared values (RMSE) for the different RBF parametrizations using $1600$ test samples. The results can be followed in Fig.~\ref{fig:E123}, where the RMSE dependence on $A$ and $C$ is shown on a contour plot, where the darkest parts correspond to the smaller errors. On the upper left subfigure the scatter plot of the true vs. predicted values is shown for the optimal case that corresponds to the parametrization $C=1.7$, $A=2.9$, while on the upper right scatter plot a 'bad' parametrization is shown with $C=3.9$, $A=2.9$ that fails to reproduce the approximable function with a 'good' accuracy. As the calculations for the silver blaze region only need one RBF network, we will use the previously determined optimal parameters with $K=10^2$ number of kernels, where the centers are distributed on a uniform grid in the range of $c_k \in [-1.7,1.7] \times [-1.7,1.7]$ with $\Delta c= 3.4/9$ resolution.
By having the RBF approximated partition function, one could derive the closed-form expression for the expectation value of $\langle n \rangle$, or one could take the numerical derivatives by, e.g., a simple forward difference scheme. In this work we choose the latter with $\Delta \mu =0.001$. The calculations are done up until the critical chemical potential at a lattice size of $N^4=10^4$, and the results can be followed in Fig.~\ref{fig:3}, where the phase quenched values (i.e., when $\mathcal{B}_p=0$) are also shown for comparison.
The results clearly show that there is no cancellation in the phase quenched theory, thus, $\langle n \rangle$ immediately grows with increasing chemical potential. In contrast, the full theory shows the expected $\mu$-independent behavior at large lattice sizes for small chemical potentials up until $\mu_c$. The results that we have obtained from the RBF approximation, again, are very similar to the ones coming from complex Langevin dynamics in \cite{25}, and it is another confirmation that the method is giving the good results.

\section{Numerical complexity}
In this section the numerical and time complexity of the method and its dependence on dimensionality and lattice size will be discussed. In general, on a cubic lattice with $N$ number of lattice sites in each dimension, the complexity of the RBF method in momentum space can be described in $\mathcal{O}(K N^D)$, where $K$ is the number of Gaussian kernels, $N$ is the number of lattice sites in one space-time direction, and $D$ is the dimensionality of the system. This general structure stays the same even if the number of lattice sites are different in the space and in the time directions, in which case one has $\mathcal{O}(K N_{space}^{D-1} N_{time})$. 
According to this, it is expected that the time it takes to calculate the different observables, such as the number density, scales linearly with the total number of lattice sites, which is very desirable when one wants to take, e.g., the thermodynamic or the continuum limits. Here, the time scaling will be discussed through the calculation of the number densities using the same system configuration that was used in the previous section, namely $m=1$ mass and $\lambda=1$ coupling strength.

To address the time complexity of the model, the number density defined in Eq.~\ref{eq:17} is calculated at $N_{\mu}=24$ chemical potential values between $\mu \in [0,1.15]$ for the full theory, i.e., the same calculation is carried out as it is shown in Fig.~\ref{fig:3}. Algorithmically, the calculation can be separated into the following steps:

1) First, the parameters of the RBF network have to be optimized by the same method that is described in the previous section. This takes a fixed time $T_{RBF}$ that does not scale with the lattice size but only with the number of kernels. 

2) The next step is the determination of $\ln Z^{rbf}$ for every chemical potential, which requires a D-dimensional nested loop (in this case $D=4$), where the $\mathcal{A}_p$ and $\mathcal{B}_p$ parameters defined in Eq.~\ref{eq:12} have to be calculated at every momenta. As $\mathcal{A}_p$ also depends on the chemical potential, it cannot be taken out from the main loop, where we go through all $\mu$, thus, these nested loops have to be done $N_{\mu}$ times. 

3) Lastly, the derivatives are calculated numerically by using a small $\Delta \mu$ step. This is done by making an extra calculation of $\ln Z^{rbf}$ for each $\mu$, so that the derivative can be approximated by using, e.g., the forward scheme as $(\ln Z^{rbf}(\mu+\Delta \mu)-\ln Z^{rbf}(\mu))/\Delta \mu$. 

To summarize, the full calculation of the number density of the full theory using the RBF approximation on $N_{\mu}=24$ points needs the determination of the partition function $2N_{\mu}$ times, where in each calculation we have a 4-dimensional nested loop ($N^4$ terms) that represents the full sum over all momentum modes. The time it takes to do the full calculation will be addressed as follows:
\begin{equation}
T=T_{RBF} + T_L,
\end{equation}
where $T_{RBF}$ represents the time it takes to train one RBF network, while $T_L$ means the time to calculate the number densities at all the given chemical potentials. In general $T_{RBF} \ll T_L$, however its actual value depends on the optimization method, number of kernels, number of training/validation/test samples, etc. On a standard notebook (LG Gram, Intel Core i5, 16GB memory) the determination of the RBF parameters for one $(K,A,c_k)$ configuration in the $K$ range we will do the comparisons ($K \sim 10^2$) is $T_{RBF}\sim 0.1$ [s]. Note that in the determination of the optimal configuration, e.g., in Fig.~\ref{fig:E123}, one needs a set of different configurations, and the full optimization time will be larger.
\begin{figure}
\centering
\includegraphics[width=\columnwidth]{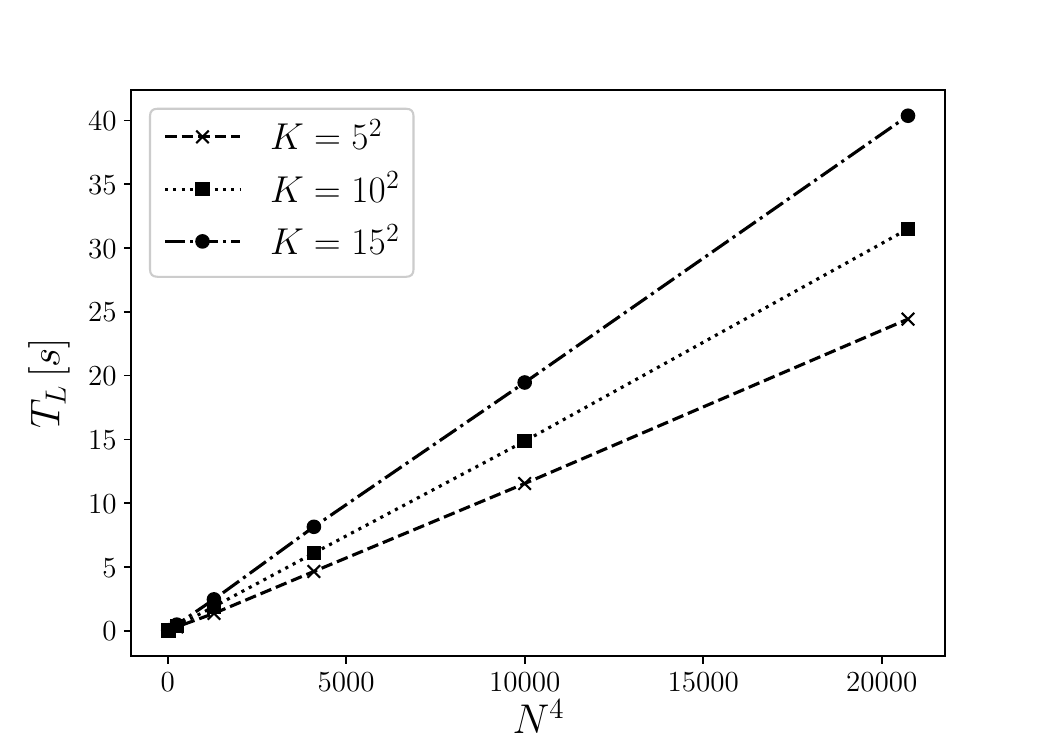}
\caption{Time complexity as a function of lattice size $N^4$ and number of Gaussian kernels $K$ in computing the number density at $N_{\mu }=24$ chemical potential values.}
\label{fig:T}
\end{figure}
To examine the scaling behavior of $T_L$, the calculations have been made by setting $K=5^2,10^2,15^2$ and $N^4=2^4,4^4,6^4,8^4,10^4,12^4$. The results can be seen in Fig.~\ref{fig:T}, where the assumed linear scaling with $K$ and $N^4$ is reproduced. According to the obtained results, the time that it takes to calculate the number densities in $24$ points (using the the optimal RBF network that is determined in Fig.~\ref{fig:E123}) is $T_L \sim 14$ [s]. 

According to the algorithm, the numerically most expensive part of the calculations is the 4-dimensional nested loops, which has to be done in every step due to the $\mu$ dependence of the $\mathcal{A}_p$ and $\mathcal{B}_p$ parameters. The time it takes to calculate one point also depends on the number of kernel functions $K$, that is also an expected behavior of the model.

\section{Conclusions}
In this paper, we have shown that the system of relativistic Bose gas at finite chemical potentials can be accurately described by RBF-approximated Euclidean path integral formalism, where the nonlinear self-interacting terms are expanded by a sum of Gaussian kernels, thus making the overall path integral solvable. Using carefully chosen symmetric centers around zero (and some corresponding width parameters), the logarithm of the partition function can be estimated in $\mathcal{O}(KN^4)$ complexity by a few percentage relative error. The RBF-approximated path integral is not constrained by the appearing imaginary terms in the Euclidean action, thus, the finite density system can be solved in the same manner as in the case of $\mu=0$. The method has been applied to the interacting complex scalar field theory, where the critical chemical potential is estimated to be $\mu_c = 1.17 \pm 0.018$ in correspondence with the results coming from complex Langevin dynamics. The silver blaze region is also addressed, where the approximate $\mu$ independent behavior of $\langle n \rangle$ on a finite lattice is well-described.

The RBF method is able to provide results in a very fast manner, e.g., the calculation regarding the silver blaze problem in Fig.~\ref{fig:3} on a $10^4$ lattice took less than a minute on a standard notebook without any optimization or using any GPU or special hardware. Due to the generality of the model, there is a possibility to extend it to include fermionic fields and non-abelian gauge theories and address problems in QCD at finite densities as well. 
On the technical side, due to its versatility, the RBF model could also be extended or formulated in other ways that could be more suitable to specific problems. 

\section{acknowledgments}
\begin{acknowledgments}
This work was supported by the Korea National Research Foundation under Grant No. 2023R1A2C300302311 and 2023K2A9A1A0609492411, and the Hungarian OTKA fund K138277.
\end{acknowledgments}

\bibliography{Main.bib}

@article{1,
   title={Functional Integrals for QCD at Nonzero Chemical Potential and Zero Density},
   volume={91},
   ISSN={1079-7114},
   url={http://dx.doi.org/10.1103/PhysRevLett.91.222001},
   DOI={10.1103/physrevlett.91.222001},
   number={22},
   journal={Physical Review Letters},
   publisher={American Physical Society (APS)},
   author={Cohen, Thomas D.},
   year={2003},
   month=nov }

@article{2,
   title={Computational Complexity and Fundamental Limitations to Fermionic Quantum Monte Carlo Simulations},
   volume={94},
   ISSN={1079-7114},
   url={http://dx.doi.org/10.1103/PhysRevLett.94.170201},
   DOI={10.1103/physrevlett.94.170201},
   number={17},
   journal={Physical Review Letters},
   publisher={American Physical Society (APS)},
   author={Troyer, Matthias and Wiese, Uwe-Jens},
   year={2005},
   month=may }

@article{R1,
   title={A new method to study lattice QCD at finite temperature and chemical potential},
   volume={534},
   ISSN={0370-2693},
   url={http://dx.doi.org/10.1016/S0370-2693(02)01583-6},
   DOI={10.1016/s0370-2693(02)01583-6},
   number={1–4},
   journal={Physics Letters B},
   publisher={Elsevier BV},
   author={Fodor, Z. and Katz, S.D.},
   year={2002},
   month=may, pages={87–92} }

@article{R2,
   title={The QCD phase diagram for small densities from imaginary chemical potential},
   volume={642},
   ISSN={0550-3213},
   url={http://dx.doi.org/10.1016/S0550-3213(02)00626-0},
   DOI={10.1016/s0550-3213(02)00626-0},
   number={1–2},
   journal={Nuclear Physics B},
   publisher={Elsevier BV},
   author={de Forcrand, Philippe and Philipsen, Owe},
   year={2002},
   month=oct, pages={290–306} }

@article{R3,
   title={Finite density QCD via an imaginary chemical potential},
   volume={67},
   ISSN={1089-4918},
   url={http://dx.doi.org/10.1103/PhysRevD.67.014505},
   DOI={10.1103/physrevd.67.014505},
   number={1},
   journal={Physical Review D},
   publisher={American Physical Society (APS)},
   author={D’Elia, Massimo and Lombardo, Maria-Paola},
   year={2003},
   month=jan }

@article{R4,
   title={The density of states method at non-zero chemical potential},
   volume={2007},
   ISSN={1029-8479},
   url={http://dx.doi.org/10.1088/1126-6708/2007/03/121},
   DOI={10.1088/1126-6708/2007/03/121},
   number={03},
   journal={Journal of High Energy Physics},
   publisher={Springer Science and Business Media LLC},
   author={Fodor, Zoltan and Katz, Sandor D and Schmidt, Christian},
   year={2007},
   month=mar, pages={121–121} }

@article{3,
   title={Stochastic quantization at finite chemical potential},
   volume={2008},
   ISSN={1029-8479},
   url={http://dx.doi.org/10.1088/1126-6708/2008/09/018},
   DOI={10.1088/1126-6708/2008/09/018},
   number={09},
   journal={Journal of High Energy Physics},
   publisher={Springer Science and Business Media LLC},
   author={Aarts, Gert and Stamatescu, Ion-Olimpiu},
   year={2008},
   month=sep, pages={018–018} }

@article{4,
   title={Lattice simulations of real-time quantum fields},
   volume={75},
   ISSN={1550-2368},
   url={http://dx.doi.org/10.1103/PhysRevD.75.045007},
   DOI={10.1103/physrevd.75.045007},
   number={4},
   journal={Physical Review D},
   publisher={American Physical Society (APS)},
   author={Berges, J. and Borsányi, Sz. and Sexty, D. and Stamatescu, I.-O.},
   year={2007},
   month=feb }

@article{5,
   title={Real-time gauge theory simulations from stochastic quantization with optimized updating},
   volume={799},
   ISSN={0550-3213},
   url={http://dx.doi.org/10.1016/j.nuclphysb.2008.01.018},
   DOI={10.1016/j.nuclphysb.2008.01.018},
   number={3},
   journal={Nuclear Physics B},
   publisher={Elsevier BV},
   author={Berges, Jürgen and Sexty, Dénes},
   year={2008},
   month=aug, pages={306–329} }

@article{7,
    author = "Carleo, Giuseppe and Cirac, Ignacio and Cranmer, Kyle and Daudet, Laurent and Schuld, Maria and Tishby, Naftali and Vogt-Maranto, Leslie and Zdeborov{\'a}, Lenka",
    title = "{Machine learning and the physical sciences}",
    eprint = "1903.10563",
    archivePrefix = "arXiv",
    primaryClass = "physics.comp-ph",
    doi = "10.1103/RevModPhys.91.045002",
    journal = "Rev. Mod. Phys.",
    volume = "91",
    number = "4",
    pages = "045002",
    year = "2019"
}

@article{8,
  title = {Estimating the Euclidean quantum propagator with deep generative modeling of Feynman paths},
  author = {Che, Yanming and Gneiting, Clemens and Nori, Franco},
  journal = {Phys. Rev. B},
  volume = {105},
  issue = {21},
  pages = {214205},
  numpages = {8},
  year = {2022},
  month = {Jun},
  publisher = {American Physical Society},
  doi = {10.1103/PhysRevB.105.214205},
  url = {https://link.aps.org/doi/10.1103/PhysRevB.105.214205}
}

@article{9,
  title = {Variational Neural-Network Ansatz for Continuum Quantum Field Theory},
  author = {Martyn, John M. and Najafi, Khadijeh and Luo, Di},
  journal = {Phys. Rev. Lett.},
  volume = {131},
  issue = {8},
  pages = {081601},
  numpages = {7},
  year = {2023},
  month = {Aug},
  publisher = {American Physical Society},
  doi = {10.1103/PhysRevLett.131.081601},
  url = {https://link.aps.org/doi/10.1103/PhysRevLett.131.081601}
}

@misc{NN1,
      title={On the solution of Euclidean path integrals with neural networks}, 
      author={Gabor Balassa},
      year={2025},
      eprint={2509.16953},
      archivePrefix={arXiv},
      primaryClass={hep-ph},
      url={https://arxiv.org/abs/2509.16953}, 
}

@article{10,
title = {Finite-density lattice QCD and sign problem: Current status and open problems},
journal = {Progress in Particle and Nuclear Physics},
volume = {127},
pages = {103991},
year = {2022},
issn = {0146-6410},
doi = {https://doi.org/10.1016/j.ppnp.2022.103991},
url = {https://www.sciencedirect.com/science/article/pii/S0146641022000497},
author = {Keitaro Nagata},
}

@article{11,
  title = {Computational Complexity and Fundamental Limitations to Fermionic Quantum Monte Carlo Simulations},
  author = {Troyer, Matthias and Wiese, Uwe-Jens},
  journal = {Phys. Rev. Lett.},
  volume = {94},
  issue = {17},
  pages = {170201},
  numpages = {4},
  year = {2005},
  month = {May},
  publisher = {American Physical Society},
  doi = {10.1103/PhysRevLett.94.170201},
  url = {https://link.aps.org/doi/10.1103/PhysRevLett.94.170201}
}

@article{12,
   title={Complex Langevin dynamics at finite chemical potential: mean field analysis in the relativistic Bose gas},
   volume={2009},
   ISSN={1029-8479},
   url={http://dx.doi.org/10.1088/1126-6708/2009/05/052},
   DOI={10.1088/1126-6708/2009/05/052},
   number={05},
   journal={Journal of High Energy Physics},
   publisher={Springer Science and Business Media LLC},
   author={Aarts, Gert},
   year={2009},
   month=may, pages={052–052} }

@article{13,
   title={Method for simulating O(N) lattice models at finite density},
   volume={75},
   ISSN={1550-2368},
   url={http://dx.doi.org/10.1103/PhysRevD.75.065012},
   DOI={10.1103/physrevd.75.065012},
   number={6},
   journal={Physical Review D},
   publisher={American Physical Society (APS)},
   author={Endres, Michael G.},
   year={2007},
   month=mar }

@article{14,
title = {Lattice study of the Silver Blaze phenomenon for a charged scalar $\phi^4$ field},
journal = {Nuclear Physics B},
volume = {869},
number = {1},
pages = {56-73},
year = {2013},
issn = {0550-3213},
doi = {https://doi.org/10.1016/j.nuclphysb.2012.12.005},
url = {https://www.sciencedirect.com/science/article/pii/S0550321312006669},
author = {Christof Gattringer and Thomas Kloiber}
}

@article{15,
    author = {Balassa, Gábor},
    title = {Fixed-energy inverse scattering with radial basis function neural networks and its application to neutron–$\alpha$ interactions},
    journal = {Progress of Theoretical and Experimental Physics},
    volume = {2023},
    number = {11},
    pages = {113A01},
    year = {2023},
    month = {10},
    issn = {2050-3911},
    doi = {10.1093/ptep/ptad131},
    url = {https://doi.org/10.1093/ptep/ptad131}
}

@article{16,
  title = {Data classification by quantum radial-basis-function networks},
  author = {Shao, Changpeng},
  journal = {Phys. Rev. A},
  volume = {102},
  issue = {4},
  pages = {042418},
  numpages = {9},
  year = {2020},
  month = {Oct},
  publisher = {American Physical Society},
  doi = {10.1103/PhysRevA.102.042418},
  url = {https://link.aps.org/doi/10.1103/PhysRevA.102.042418}
}

@misc{RBF,
      title={Neural network approximation of Euclidean path integrals and its application for the $\phi^4$ theory in 1+1 dimensions}, 
      author={Gabor Balassa},
      year={2025},
      eprint={2509.18785},
      archivePrefix={arXiv},
      primaryClass={hep-ph},
      url={https://arxiv.org/abs/2509.18785}, 
}

@article{17,
   title={Cooling stochastic quantization with colored noise},
   volume={96},
   ISSN={2470-0029},
   url={http://dx.doi.org/10.1103/PhysRevD.96.114505},
   DOI={10.1103/physrevd.96.114505},
   number={11},
   journal={Physical Review D},
   publisher={American Physical Society (APS)},
   author={Pawlowski, Jan M. and Stamatescu, Ion-Olimpiu and Ziegler, Felix P. G.},
   year={2017},
   month=dec }

@article{18,
   title={Improved lattice measurement of the critical coupling in $phi_2^4$ theory},
   volume={79},
   ISSN={1550-2368},
   url={http://dx.doi.org/10.1103/PhysRevD.79.056008},
   DOI={10.1103/physrevd.79.056008},
   number={5},
   journal={Physical Review D},
   publisher={American Physical Society (APS)},
   author={Schaich, David and Loinaz, Will},
   year={2009},
   month=mar }

@article{19,
    author = "Arnold, Peter Brockway and Espinosa, Olivier",
    title = "{The Effective potential and first order phase transitions: Beyond leading-order}",
    eprint = "hep-ph/9212235",
    archivePrefix = "arXiv",
    reportNumber = "UW-PT-92-18, USM-TH-60",
    doi = "10.1103/PhysRevD.47.3546",
    journal = "Phys. Rev. D",
    volume = "47",
    pages = "3546",
    year = "1993",
    note = "[Erratum: Phys.Rev.D 50, 6662 (1994)]"
}

@article{20,
    author = "Callaway, David J. E. and Maloof, David J.",
    title = "{Effective Potential of Lattice $\phi^4$ Theory}",
    reportNumber = "ANL-HEP-PR-82-19",
    doi = "10.1103/PhysRevD.27.406",
    journal = "Phys. Rev. D",
    volume = "27",
    pages = "406",
    year = "1983"
}

@misc{21,
      title={Cosmological Phase Transitions}, 
      author={Norbert Straumann},
      year={2004},
      eprint={astro-ph/0409042},
      archivePrefix={arXiv},
      primaryClass={astro-ph},
      url={https://arxiv.org/abs/astro-ph/0409042}, 
}

@article{22,
   title={Convex effective potential of O(N)-symmetricø4 theory for large N},
   volume={479},
   ISSN={0550-3213},
   url={http://dx.doi.org/10.1016/0550-3213(96)00403-8},
   DOI={10.1016/0550-3213(96)00403-8},
   number={3},
   journal={Nuclear Physics B},
   publisher={Elsevier BV},
   author={Mukaida, Hisamitsu and Shimada, Yujiro},
   year={1996},
   month=nov, pages={663–682} }

@article{23,
   title={Spontaneous symmetry breaking and linear effective potentials},
   volume={86},
   ISSN={1550-2368},
   url={http://dx.doi.org/10.1103/PhysRevD.86.025028},
   DOI={10.1103/physrevd.86.025028},
   number={2},
   journal={Physical Review D},
   publisher={American Physical Society (APS)},
   author={Alexandre, J.},
   year={2012},
   month=jul }

@article{25,
   title={Can Stochastic Quantization Evade the Sign Problem? The Relativistic Bose Gas at Finite Chemical Potential},
   volume={102},
   ISSN={1079-7114},
   url={http://dx.doi.org/10.1103/PhysRevLett.102.131601},
   DOI={10.1103/physrevlett.102.131601},
   number={13},
   journal={Physical Review Letters},
   publisher={American Physical Society (APS)},
   author={Aarts, Gert},
   year={2009},
   month=apr }

@ARTICLE{AIC,
  author={Akaike, H.},
  journal={IEEE Transactions on Automatic Control}, 
  title={A new look at the statistical model identification}, 
  year={1974},
  volume={19},
  number={6},
  pages={716-723},
  doi={10.1109/TAC.1974.1100705}}

\end{document}